\documentstyle [12pt,aaspp4]{article}
\newcommand{\simgt}{\lower.5ex\hbox{$\; \buildrel > \over \sim \;$}}
\newcommand{\simlt}{\lower.5ex\hbox{$\; \buildrel < \over \sim \;$}}

\def\pppm{\rm P^3M}
\newcommand{\citet}[1] {\cite{#1}}
\newcommand{\citep}[1] {(\cite{#1})}
\slugcomment{SHAO/MPA-003}
\begin{document}
\title{A new test for the stable clustering hypothesis}

\author{Y.P. Jing$^{1,2}$}  
\affil{$ ^1$
Shanghai Astronomical Observatory, the Partner Group of MPI f\"ur
Astrophysik, Nandan Road 80, Shanghai 200030, China}
\affil {$^2$ National Astronomical Observatories, Chinese Academy of
Sciences, Beijing 100012, China}
\affil{ypjing@center.shao.ac.cn}

\received{2001 January}
\accepted{2001 ???}

\begin{abstract}
The stable clustering hypothesis is a fundamental assumption about the
nonlinear clustering of matter in cosmology. It states that the mean
physical separation of particles is a constant on sufficiently small
scales. While many authors have attempted to test the hypothesis with
cosmological N-body simulations, no consensus has been reached on
whether and where the hypothesis is valid, because of the limited
dynamical range that this type of simulations can achieve. In this {\it
Letter}, we propose to test the hypothesis with high resolution halo
simulations, since the individual halo simulations can resolve much
better the fine structures of the halos and since almost all pairs of
particles with small separations are presumed to be inside virialized
halos. We calculated the mean pair velocity for 14 high resolution
halos of $\sim 1$ million particles in a low-density flat cold dark
matter model. The result agrees very well with the stable clustering
prediction within the measurement uncertainty $\sim 30\%$ over a large
range of scales where the overdensity is $10^3$ to $10^6$. The
accuracy of the test can be improved to $\sim 10\%$ if some 100 halos
with a similar resolution are analyzed.

\end{abstract} 


\section{Introduction}
The stable clustering hypothesis is a basic assumption in cosmology
about the nonlinear clustering of cosmic matter
(\cite{Peebles1980}). It states that the mean physical separations $r$
of particles in highly non-linear regions do not change with the
expansion of the Universe. In terms of the scale factor $a$ and the
comoving coordinates ${\bf x}$, the hypothesis equivalently states
that the relative pair velocity in the comoving coordinates $v=a
\dot{x}$ cancels out the Hubble flow $\dot{a}x$, i.e.
\begin{equation}
v=-\dot{a}x=-Hr\,.  
\label{eq1}
\end{equation}
Based on this assumption plus self-similarity, it is 
predicted that 1) the power-law index
of the two-point correlation function $\xi(r)$ on sufficiently small
scales is related to the power-law index $n$ of the linear power spectrum $P(k) \propto
k^{n}$ as $\xi(r) \propto r^{-3(3+n)/(5+n)}$, and 2) the N-point
correlation functions $\zeta_{N}$ scale with $\xi$ as
$\zeta_{N}\propto \xi^{N-1}$ (\cite{P74}, \cite{DP77},
\cite{GR75}, \cite{Peebles1980},  \cite{bs89}, \cite{Suginohara1991}, \cite{Suto1993}, 
\cite{Jain1997}).  Based on the
first prediction for clustering on small scale and the linear theory
prediction for large scale, a phenomenological fitting formula has
been found for the nonlinear power spectrum (\cite{Hamilton1991},
\cite{JMW1995}, \cite{PD1996}, \cite{Ma1998}, \cite{Suginohara2001}). 
This fitting formula has been applied to study a wide range of
non-linear clustering problems, notably the gravitational lensing and
the non-linear clustering of galaxies.

There have been many attempts to verify the stable clustering
hypothesis with numerical N-body simulations. In their seminal work,
Efstathiou et al. (\cite{efs}) tested the hypothesis [Eq.(\ref{eq1})]
and its two predictions with a set of scale-free N-body simulations of
$32^3$ particles. They concluded that their simulation results are
consistent with the hypothesis, but it is not difficult to imagine
that the accuracy they could achieve is quite limited for the limited
resolution of this first generation of simulations.  Subsequently,
many improved simulations with $64^3$ to $128^3$ particles have been
used to solve this problem (e.g. \cite{Suginohara1991},
\cite{Suto1993}, \cite{cbh1995}, \cite{munshietal1997}, and
\cite{Jain1997} and references therein). As shown by Jain (1997; his
Fig.2, Fig.3 and Fig.8), his simulation results of the mean pair
velocity have just approached the stable clustering condition
[Eq.(\ref{eq1})] at the smallest scale that his simulations of $\sim
128^3$ particles can resolve. Thus, it remains an open question if the
stable clustering holds for a range of scales
(\cite{munshietal1997}). One point clear from these previous studies
is that only in high overdensity regions $\Delta \rho/\rho\gg 100$ is
the stable clustering assumption possibly valid. One needs simulations
which have much higher resolutions than those Jain analyzed, in order
to convincingly test the stable clustering.

It is well known that close pairs of particles reside inside virialized
halos (e.g. \cite{mcs77}, \cite{Peebles1980}). This is
the reason why many people have  recently constructed
successful models for non-linear statistics based on the
simulation result of the halo density profile and 
the halo mass function (e.g. \cite{mjb},
\cite{sj97}, Ma \& Fry 2000a,b,c, \cite{uros},
\cite{CHM00}, \cite{jp},
\cite{halo}, \cite{shethetal2000}). In particular,
Ma \& Fry (2000b) checked the stable clustering assumption using such
an approach. They derived the mean pair velocity by inserting the halo
model two-point correlation function into the pair conservation
equation (e.g. \cite{Peebles1980}), and found that the stable clustering is
valid only when the halo mass function and the halo density profile
obey specific relations.  Considering the fact that both models of the
halo density profile and the halo mass function have
considerable uncertainties and scatters 
(\cite{jing2000}, \cite{JS00}, 
\cite{st99}, \cite{J00}), they concluded
that it is yet unclear if the stable clustering can be realized in
practice.

In this {\it Letter}, we propose to use high-resolution halo
simulations to test the stable clustering hypothesis by examining the
mean pair velocity within virialized halos.  The validity of this
approach underlies the above-mentioned fact that almost all close
pairs of galaxies reside inside virialized halos. If the mean pair
velocity within halos agrees with the stable clustering hypothesis,
one can sufficiently conclude that the stable clustering is
established, since the mean pair velocity in general is a
$N_{pair}^h(r,M) n(M)$ weighted average of the mean pair velocity
within halos, where $N_{pair}^h(r,M)$ is the mean pair number of
separation $r$ inside a halo of mass $M$, and $n(M)$ is the halo
number density per unit halo mass. The advantage of this approach over the
conventional approach which uses cosmological N-body simulations
(e.g. Jain 1997) is that the individual halos can be simulated with a
much better resolution. Here we use fourteen high-resolution halos to
make such an analysis. Twelve of these halos were used for studying
density profiles of dark matter halos (Jing \& Suto 2000) and the
other two were simulated after the work of Jing \& Suto (2000).

\section{Simulation data}

Fourteen dark matter halos in total were selected from our one
cosmological P$^3$M N-body simulation of $256^3$ particles in a
$(100h^{-1}{\rm Mpc})^3$ cube (\cite{JS1998}, \cite{Jing1998}).  The
model for this simulation is the standard Lambda cold dark matter
model which has the density parameter $\Omega_0=0.3$, the cosmological
constant $\lambda_0=0.7$, the Hubble constant $h=0.7$ and the power
spectrum normalization $\sigma_8=1.0$. These model parameters are
consistent with the current observations of the cluster abundance, of
the cosmic background radiation anisotropies, of high-redshift
supernova luminosity, and of many others. Of the fourteen halos, five,
five, and four have mass scales of clusters, groups and galaxies
respectively.

These halos are re-simulated with high resolution using the multiple
mass method (e.g. \cite{kw93}) and our nested-grid $\pppm$ code
(\cite{JS00}).  To minimize the contamination of the coarse particles
on the halo properties within the virial radius at $z=0$, $r_{\rm
vir}$, we trace back the particles within $3r_{\rm vir}$ of each halo
to their initial conditions at redshift $z=72$. The virial radius
$r_{\rm vir}$ is defined such that the spherical overdensity inside is
$\sim 18\pi^2 \Omega_0^{0.4} \sim 110$ times the critical density
$\rho_{\rm crit}(z=0)$ (\cite{Kitayama1997}, \cite{bn1998}). Typically
$ 1.5\times 10^6$ fine particles are placed in the high resolution
region, and $ 0.7\times 10^6$ coarse particles, whose mass increases
monotonically with the distance from the high-resolution region, are
placed outside the high resolution region to model the tidal force
effect. The initial conditions for these particles are produced with
the same random initial field as the cosmological N-body simulation
but with the shorter wavelength perturbations added that the
cosmological N-body simulation missed. The simulations are evolved by
our nested-grid $\pppm$ code which was designed to simulate
high-resolution halos.  The force resolution is typically $0.004
r_{vir}$.  At last about $(0.5 - 1)\times 10^6$ particles end up
within $r_{\rm vir}$ of each halo.  For more details about the
simulation, we refer readers to Jing \& Suto (2000).

\section{The mean pair velocity}
We calculate the mean pair velocity $v_r(r)$ by averaging the relative
velocity of all pairs of particles within the virial
radius. Separation bins are taken from $0.01r_{vir}$ to $r_{vir}$ with
$\Delta \lg r =0.05$. Since one halo has $(0.5 - 1)\times 10^6$
particles, there are more than $10^{11}$ pairs in total in each halo,
and more than $10^{6}$ pairs within the single innermost bin. The
high number of pairs is very crucial for investigating $v_r(r)$ at the
smallest separation. This is because relative to the stable clustering
prediction $v_r=-0.01H_0r_{vir}$ at $r=0.01r_{vir}$, the random
motion of particles (i.e. the noise for determining $v_r$) is very
large which, in terms of the one-dimensional velocity
dispersion, is about $5H_0r_{vir}$. In order to suppress the noise,
more than $500^2=2.5\times 10^5$ particle pairs are needed for the
innermost bin. Only high-resolution halo simulations like those used
in this paper can easily meet this requirement. In contrast,
cosmological N-body simulations would need many more than $128^3$
particles to study $v_r(r)$ at a similar scale.

We find that the mean pair velocity $v_r(r)$ varies significantly from
one halo to another. In Figure 1, we show $v_r(r)/H_0 r$ for the five
cluster halos as a function of $r$. The halos are labeled in the same
way as Jing \& Suto (2000) with the newly added halo labeled as
CL5. Although the measured values of $v_r(r)/H_0 r$ are around the
stable clustering prediction $-1$, there is significant difference in
this quantity between one halo and another. Some halo has $v_r(r)/H_0
r$ systematically higher than $-1$ (CL3), and some lower than $-1$ (CL1 \&
CL2). We have examined whether $v_r(r)/H_0 r$ is correlated with the
dynamical state and the density profiles of halos. From Jing \& Suto
(2000), we know that CL1 has the smallest concentration parameter and
CL2 has the largest among the first four halos, while they have
similar $v_r(r)/H_0 r$. Therefore there seems to be no apparent
correlation between the density profiles and $v_r(r)/H_0 r$, which is
also supported by the results of galaxy and group halos (Figure
2). The large variation of $v_r(r)/H_0 r$ from halo to halo might
reflect the fact that the merging and relaxation processes are
still taking place within the halos, which is not unexpected in the
hierarchical clustering scenario, and the value of $v_r(r)/H_0 r$
depends on a fine balance of the both processes within a halo.

For testing the stable clustering hypothesis, the important quantity
is not the $v_r(r)/H_0 r$ of an individual halo, but the averaged
value of $v_r(r)/H_0 r$ over halos of a given mass. Figure 2 shows the
averaged $v_r(r)/H_0 r$ for the cluster (CL), group (GR) and galaxy
(GX) halos (note that the vertical axis is different from that of
Figure 1). The measurement has a typical uncertainty $0.6\sim 1.0$
(For clarity, the error bars are plotted for the cluster halos
only). As the figure shows, the result of each set of halos is
consistent with the stable clustering hypothesis within the
measurement uncertainty.

Similarities have been found for the halos of different mass after the
length (and correspondingly other quantities) is scaled to the halo
size. Navarro et al. (1996) proposed a universal density profile for
halos, and Klypin et al (1999) and Moore et al. (1999) showed that the
number distributions of subhalos (after scaling to the circular
velocity) are very similar between galaxy and cluster halos.  Although
there is some systematic difference between halos of different
mass even after scaling (e.g. the density profiles, Jing \& Suto
2000), the small difference might not be important for the current
work.  This is reflected in Figure 2 which shows no dependence of
$v_r(r)/H_0 r$ on the halo mass. As a further step, we average
$v_r(r)/H_0 r$ over all 14 halos at each $r/r_{vir}$-bin, and the
results are presented in Figure 3. It shows that the averaged result
is well consistent with the stable clustering prediction within a
typical uncertainty $\sim 30\%$. According to the density profiles,
the typical density contrast at $r=(0.01\sim 0.1)r_{vir}$ is $10^3$ to
$10^6$.

\section{Conclusion and future work}
We have made a first attempt to investigate the validity of the
stable clustering hypothesis using high-resolution halo
simulations. We have obtained the following conclusions.
\begin{itemize}
\item The mean pair velocity within a single halo fluctuates strongly
around the stable clustering prediction. This may indicate that the
merging and relaxation processes inside halos have not fully
completed, as should be expected for hierarchical clustering scenarios.
\item The mean pair velocity averaged over four or five halos of
a similar mass is consistent with the stable clustering prediction
with an uncertainty of $\sim 60\%$. No dependence on halo mass
has been seen for the quantity $v_r(r)/H_0 r$.
\item Based on the similarities exhibited by halos of different mass,
we averaged the $v_r(r)/H_0 r$ over the fourteen halos at each
$r/r_{vir}$-bin. The result is in very good agreement with the stable
clustering assumption with an uncertainty of $\sim 30\%$ over a large
range of the scales $r$, thus lending strong support for the stable clustering
hypothesis.
\end{itemize}

It is important to emphasize again that the stable clustering
condition (1) is not generally satisfied within one single
virialized halo, but it can be achieved through averaging over many
halos.  Although there is clear evidence for the similarities among
halos of different mass, we are still concerned of any dependence of
$v_r(r)/H_0 r$ on the halo mass. We have started a program to simulate
50 halos at each mass range of clusters, groups, and galaxies, 150
halos in total. This will reduce the uncertainty of $v_r(r)/H_0 r$ to
$\sim 15\%$ for each mass range and will show the dependence of
$v_r(r)/H_0 r$ on the halo mass, if any.

\acknowledgments 
The author would like to thank Gerhard B\"orner and
Houjun Mo for helpful comments, and Yasushi Suto for his collaboration
on the simulation sample.  The research is supported in part by the
One-Hundred-Talent Program, by NKBRSF(G19990754) and by NSFC.
Numerical computations were carried out on VPP300/16R and VX/4R at
ADAC (the Astronomical Data Analysis Center) of the National
Astronomical Observatory, Japan, as well as at RESCEU (Research Center
for the Early Universe, University of Tokyo) and KEK (High Energy
Accelerator Research Organization, Japan). Several helpful comments
from the referee, S. Colombi, are acknowledged.



\newpage

\begin{figure}
\epsscale{1.0} \plotone{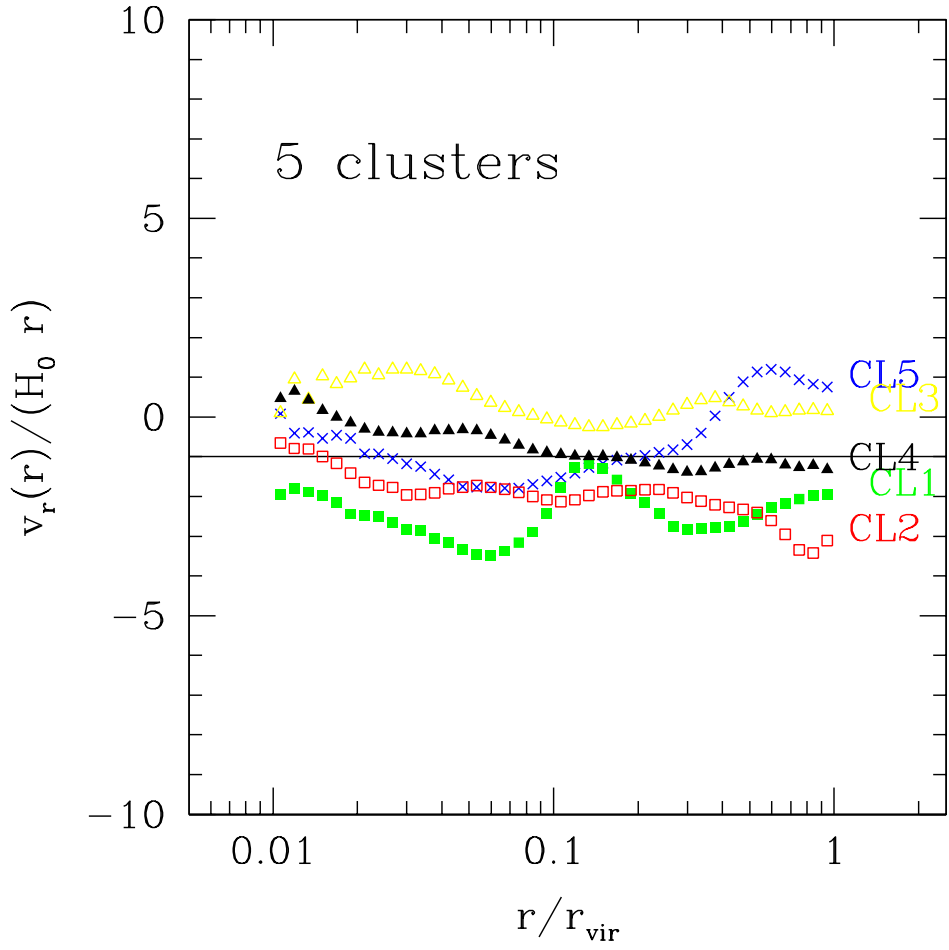}
\caption{The mean peculiar velocity, in units of $H_0r$,  of particles within cluster halos. }
\end{figure}
\begin{figure}
\epsscale{1.0} \plotone{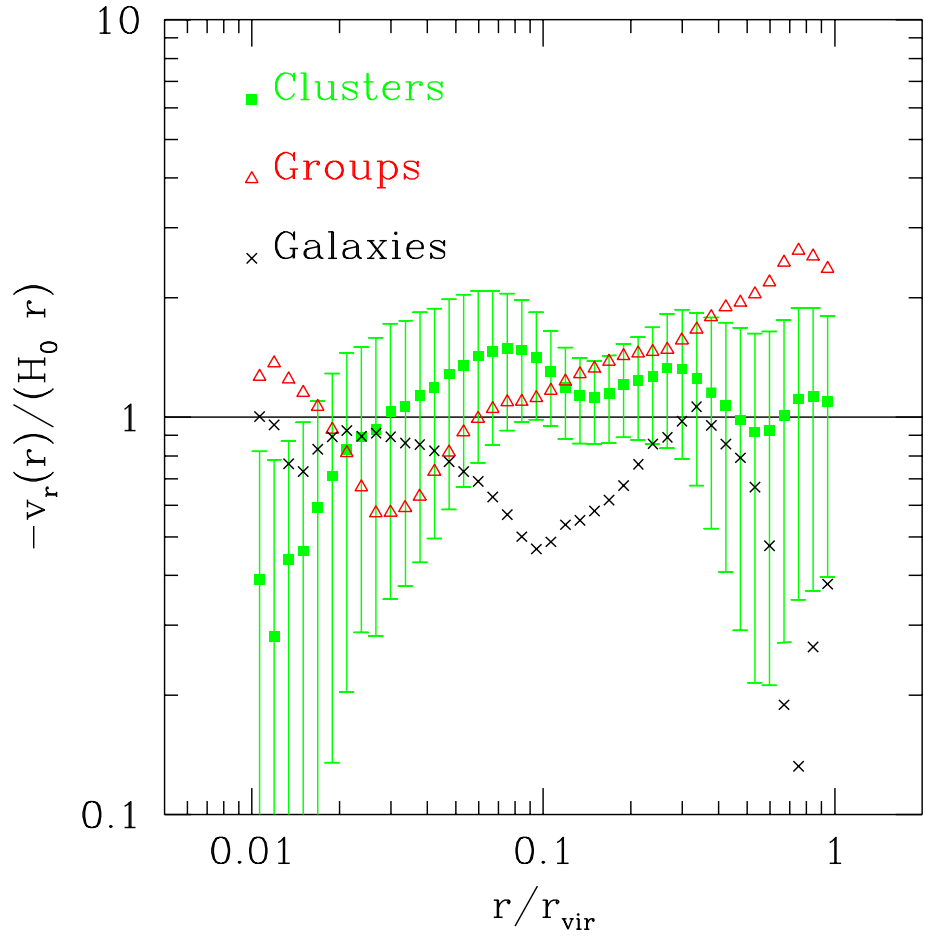}
\caption{The mean peculiar velocity, in units of $H_0r$,  of particles averaged over four to five halos at cluster, group or galaxy masses. For clarity, errorbars are plotted for the cluster halos only. }
\end{figure}

\begin{figure}
\epsscale{1.0} \plotone{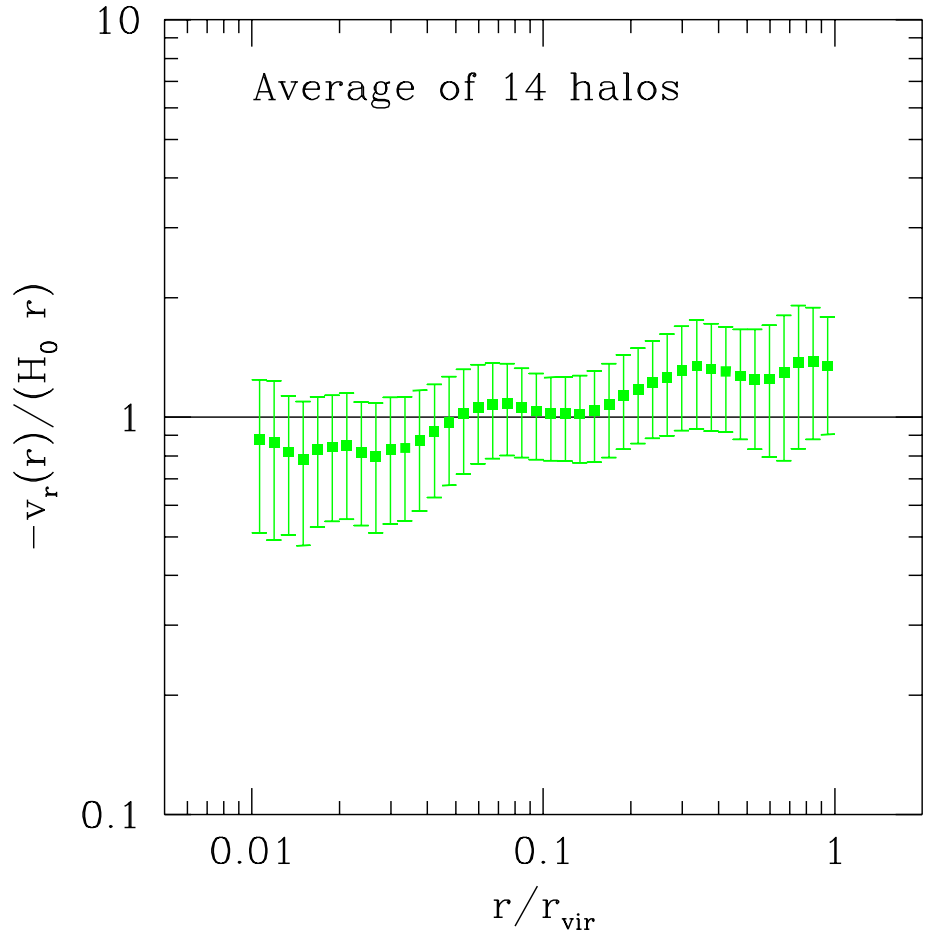}
\caption{The mean peculiar velocity, in units of $H_0r$,  of particles
 averaged over the 14 halos. }
\end{figure}

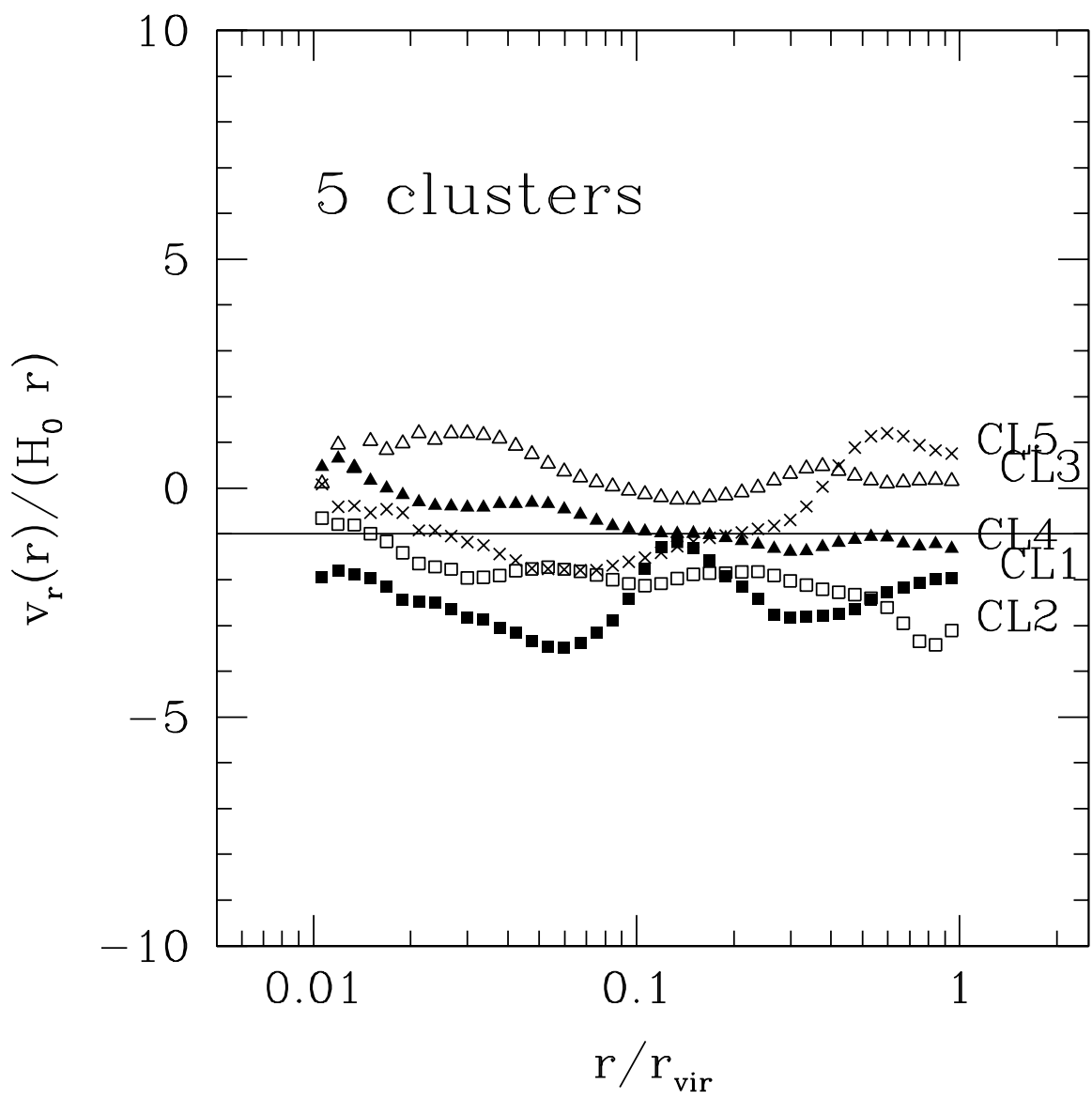
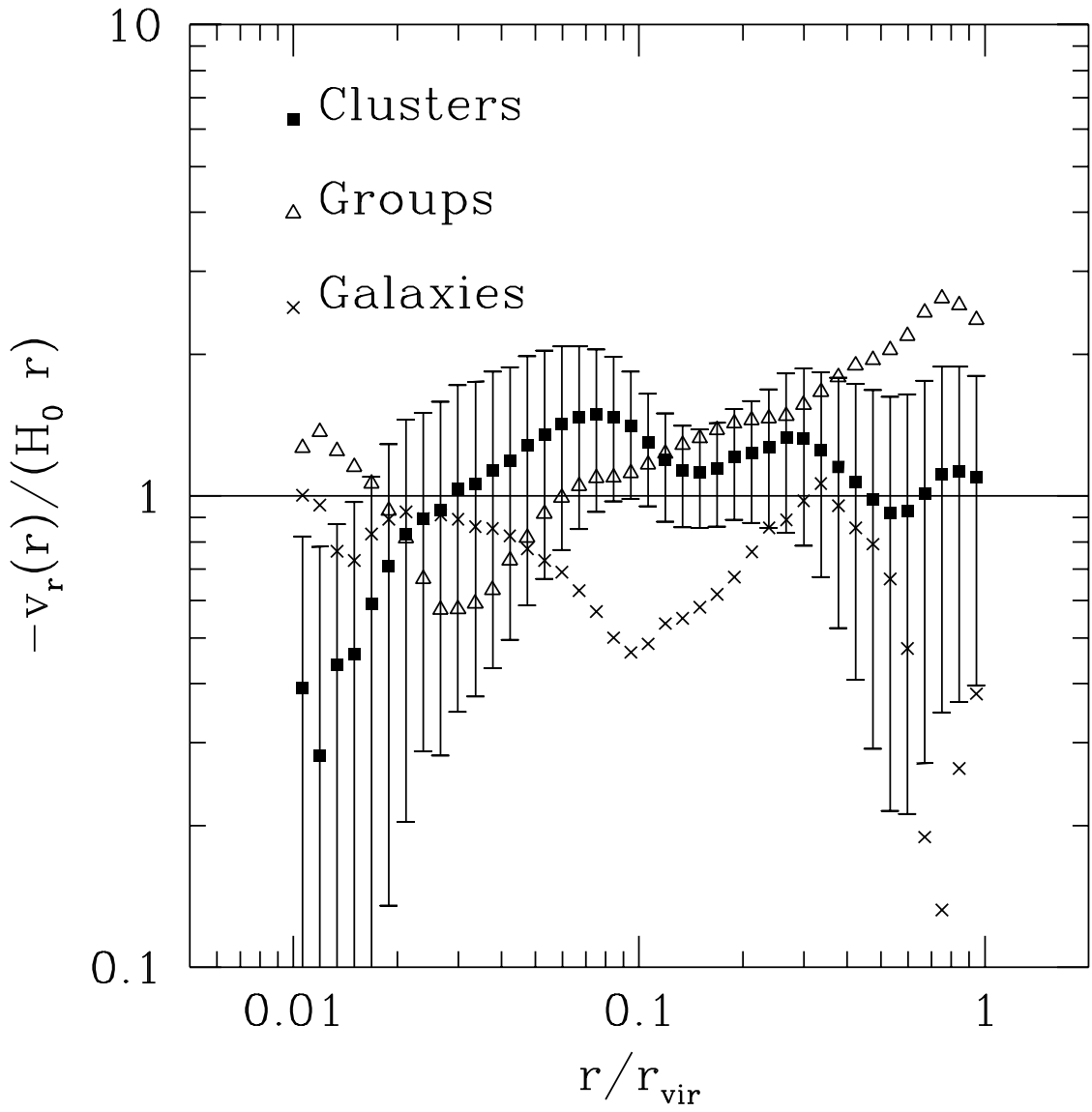
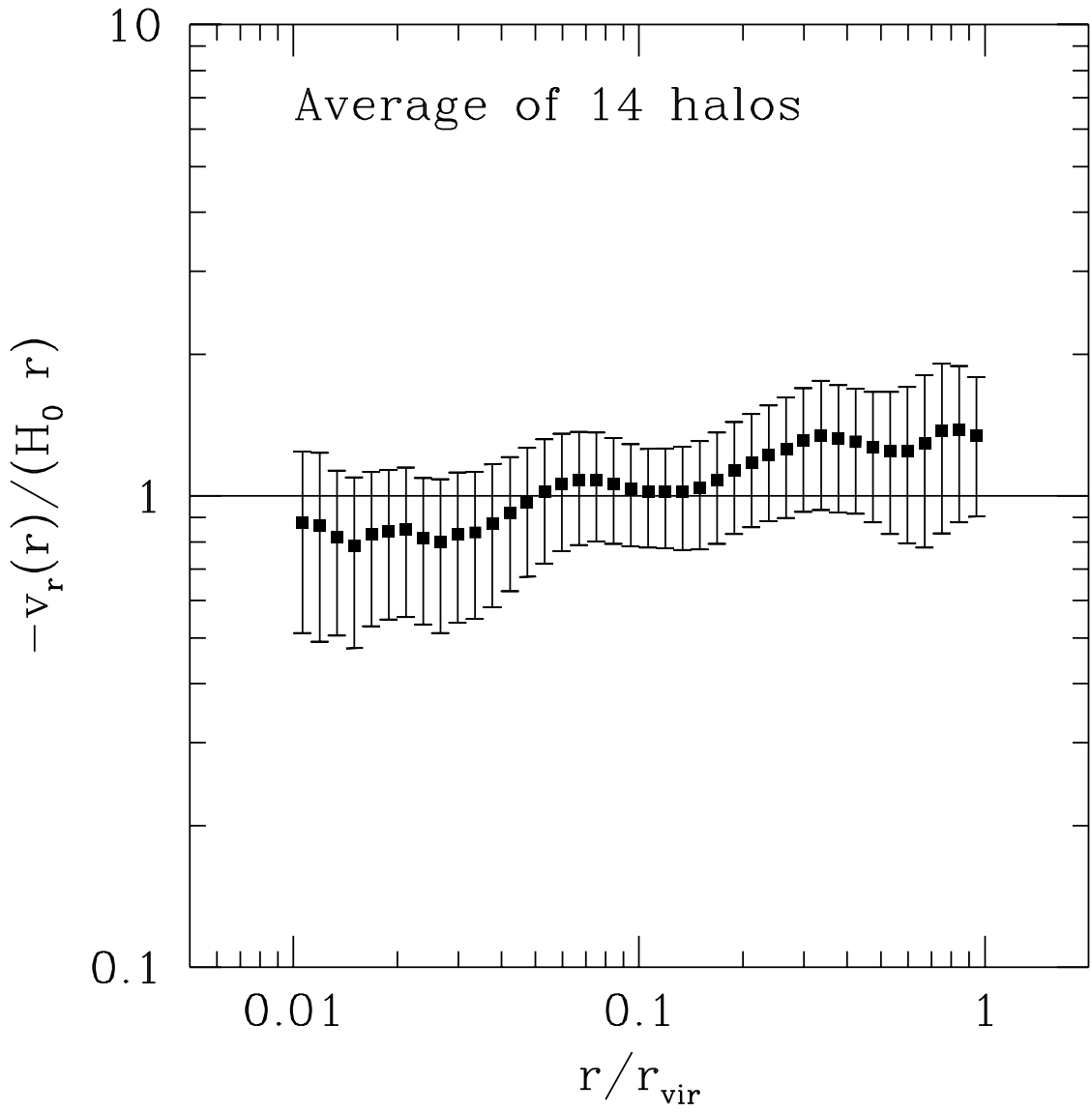
\end{document}